\documentclass[aps,prx,twocolumn,groupedaddress,showpacs,floatfix,amsmath,amssymb]{revtex4}
\usepackage{graphicx,amsmath,amssymb}
\usepackage{color}
\begin{document}
\preprint{}
\title{ 
Quantification of the volume-fraction reduction of sheared fragile glass-forming liquids and its impact on rheology
}

\author{Akira Furukawa}
\email{furu@iis.u-tokyo.ac.jp}
\affiliation{Institute of Industrial Science, University of Tokyo, Meguro-ku, 
Tokyo 153-8505, Japan. }
\date{\today}
\begin{abstract}
This study determines the volume-fraction reduction of sheared fragile glass-forming liquids.  We consider a group of hypothetical systems that consist of particles with anisotropic particle-size modulations yet have almost the same average particle configuration as actual systems under shear flow. Our molecular dynamics (MD) simulations demonstrate that one specific hypothetical system can reproduce the relaxation dynamics of an actual sheared system, and we identify the shear-flow effect on the particle size with anisotropic size-modulation of this specific system. Then, based on the determination of the particle size and the resultant volume fraction, we rationalize how slight decreases in the volume fraction significantly reduce the viscosity snf provide a nonlinear constitutive equation. Notably, the obtained rheological predictions, including the crossover shear rate from Newtonian to non-Newtonian behavior, can be expressed only in terms of experimental observables, showing a good agreement with the MD simulation results. Our perspective on the volume fraction under shear flow may provide new insights into the conventional concept of free-volume.
\end{abstract}

\maketitle

\section{Introduction}

 Shear thinning is one of the most ubiquitous non-Newtonian flow behaviors in glassy materials \cite{Yamamoto-Onuki,Berthier-Barrat,Varnik,Shi-Falk,Lemaitre,FurukawaS1,Webb-Dingwell,Kato-Kawamura-Inoue-Chen,Lu-Ravichandran-Johonson,Besseling-Isa-Ballesta-Petekidis-Cates-Poon,Liu-NagelB,LarsonB,dynamic_heterogeneityB,VoightmannR,Reyes-Sahimi,Zhu-Xia-Aitken-Sen,Chen-MaR}. When an imposed shear rate  $\dot\gamma$ is smaller than the crossover value $\dot\gamma_{\rm c}$, the shear viscosity $\hat\eta$ and the structural relaxation time $\hat\tau_\alpha$ under a given flow are the same as those at equilibrium ($\dot\gamma=0$), $\eta^{\rm (eq)}$ and $\tau_\alpha^{\rm (eq)}$, respectively. 
In contrast, when $\dot\gamma > \dot\gamma_{\rm c}$, $\hat\eta$ and $\hat\tau_\alpha$ decrease significantly as $\dot\gamma$ increases. This nonlinear flow response usually causes more complex phenomena, such as shear banding and fracture, drastically altering the mechanical properties. Thus, understanding and controlling the shear thinning properties are of particular importance in the design of processing of glassy materials; however, there is still no general consensus regarding the underlying mechanism of shear thinning. 

Among many attempts (see papers \cite{Spaepen,Taub-Spaepen,Argon,Falk-Langer,Fuchs-Cates,Brader-Cates-Fuchs,Miyazaki-Reichman,SGR,Otsuki-Sasa,Lubchenko,Trond-Tanaka,Ghosh-Schweizer,Furukawa-Tanaka1,Furukawa-Tanaka2,Yamaguchi,Lemaitre2,Langer,FurukawaS3,FurukawaS4} and the references therein) to understand the mechanism of shear thinning, the free-volume model is intuitively appealing,  and thus, has been extensively investigated. Nevertheless, the physical substance of ``free volume'' is still unclear, and we do not have enough quantitative pieces of knowledge to determine how and to what extent the free volume (volume fraction) increases (decreases) in an external flow field.

For many fragile glass-formers under equilibrium conditions, it is known that the density $n$ and the temperature $T$ are not independent parameters, whereas their combined variable determines physical states, i.e., the so-called (power law) density scaling \cite{RolandR,PaluchB,Alba,Schroder,Coslovich-Roland,Pedersen-Schroder-Dyre,DyreR,Sengupta}. The density scaling is naively interpreted as follows: increasing (decreasing) the temperature increases (decreases) thermal fluctuations and overlap between neighboring particles, which results in an effective decrease (increase) in the particle size and the volume fraction. In the sense of the density scaling, determining the effective volume fraction by setting the temperature and the (number or mass) density and then evaluating how the effective volume fraction changes by varying these parameters are fundamental. 
Now we may ask how such a physical picture for equilibrium liquids is modified for nonequilibrium liquids under an external flow field.  In an applied flow field, the interaction potential remains unchanged, while the average structural configuration is anisotropically distorted according to the flow symmetry. This distortion subsequently modifies the overlap properties from those at equilibrium which can influence determining the effective particle size. In other words, the shear rate $\dot\gamma$, in addition to the temperature $T$ and the density $n$, may serve as an extra parameter to control the effective particle size.

The present study addresses these issues with the aid of molecular dynamics (MD) simulations of model fragile glass-formers.  Attempts to quantitatively estimate the free volume or the effective volume fraction by directly analyzing {\it actual} sheared systems have not yet succeeded. Instead, in this study, we consider a group of {\it hypothetical} liquid systems with different particle sizes but with the same two-body pair correlation function as an actual sheared system. We find one such hypothetical system that can reproduce the relaxation dynamics of the actual sheared system. Then, we determine the effective particle size of the actual system by identifying it with the particle size of the specific hypothetical system. 
The detailed approach is as follows. 
(i) During the structural relaxation period ${\hat \tau}_\alpha$, the particle configurations, on average, are preserved. 
Thus, in a flow field with shear rate $\dot\gamma$ the average structure undergoes shear deformation with a strain of $\gamma=\dot\gamma {\hat \tau}_\alpha$.  
The two-body pair correlation function describes the extension (compression) of the particle configurations along the extension (compression) axis by $\gamma/2$ relative to the equilibrium state. 
However, we notice that the two-body pair correlation function can be interpreted differently. As shown in Sec. II, the particle configuration under shear flow can be reproduced by hypothetical anisotropic distortion operations applied to particles: we can prepare a group of hypothetical systems that share two-body pair correlation functions that are almost identical to that of the actual sheared system. 
(ii) The hypothetical distortion operation reproducing the steady structure under shear flow is described by two parameters: the degree of shear distortion and the particle overlap. There is arbitrariness in setting these parameters.  
Nevertheless, as shown in Sec. III B, a specific distortion operation can even reproduce the relaxation dynamics of the actual sheared system, whereby such arbitrariness can be removed. 
By identifying this specific operation with the actual shear-flow effect controlling the effective volume fraction, we provide a prescription for quantitatively determining the volume-fraction reduction in sheared fragile glass-forming liquids. 
The effective particle size along the extension axis of an externally applied flow field is the invariant reference, while in other directions it anisotropically decreases. 
(iii) Furthermore, in Sec. III C, by incorporating the volume fraction under shear flow, determined above, into the Doolittle equation of the structural relaxation time, we give a nonlinear constitutive equation. 
This constitutive equation can explain how a slight decrease (increase) in the volume fraction (free volume) significantly reduces the viscosity with high predictability. 
Although we previously derived similar rheological predictions based on more heuristic arguments in Refs. \cite{FurukawaS3,FurukawaS4}, here, we provide a detailed numerical investigation of the physical origin of the shear-induced reduction of the volume fraction or enhancement of the free volume.

\section{Theoretical Background}

In this section, we first provide a detailed explanation of the theoretical background of our study, whose validity is examined in the following section using MD simulations. A binary particle system is employed in our MD simulations to prevent crystallization, whereas, in this section, a monodisperse system is assumed for the simplicity of the expressions. The generality of the discussion presented below is not lost under this assumption.  

We assume that the constituent particles interact via the following inverse power law (IPL) potential: 
\begin{eqnarray}
U^{(A)}(r)=\epsilon \biggl(\dfrac{\sigma}{r}\biggr)^{\zeta},  \label{potentialA}
\end{eqnarray}
where $r$ is the distance between two particles and $\zeta$ is an exponent that is  sufficiently larger than 1. 
Hereafter, a system whose constituent particles interact via $U^{(A)}(r)$ is referred to as an $A$-system. 

\subsection{The effective volume fraction}

In typical simulation studies of liquids, the thermal energy scale is comparable to $\epsilon$. For such a case, when $\zeta\gg 1$, neighboring particles are strongly prevented from getting closer than a distance $\sigma$ to each other. 
Therefore, in most literature, $\sigma$ is conventionally set to the (soft) core or the particle size. 
In this setting, the particle volume and the volume fraction are simply given as 
\begin{eqnarray}
 v^{(0)} = \dfrac{1}{6}\pi \sigma^3, 
\end{eqnarray}
and 
\begin{eqnarray}
\varphi=\dfrac{N v^{(0)}}{V}=nv^{(0)},  \label{simple_volume}
\end{eqnarray} 
respectively. 
Here, $N$ is the total number of particles, $V$ is the system volume, and the particle number density is denoted as $n(=N/V)$. 

Although the above setting of the volume fraction is simple, the nature of particle packings intrinsically depends on the temperature. 
That is, as stated in the introduction, increasing (decreasing) temperature increases (decreases) thermal fluctuations, and therefore, even at a fixed $n$, the overlap between neighboring particles is enhanced (reduced), resulting in an effective decrease (increase) in the particle size and the volume fraction. Such a competing effect between repulsive interparticle interactions and thermal fluctuations is incorporated in the effective volume fraction as follows.  
In the present system with the IPL potential Eq. (\ref{potentialA}), the physical state is characterized by a scaling variable \cite{Hiwatari,Broughton,Hoover}, 
\begin{eqnarray}
\phi = 
n \sigma^3 \biggl(\dfrac{\epsilon}{T}\biggr)^{3/\zeta}, \label{effective_volume}
\end{eqnarray}  
where the temperature $T$ is measured in units of the Boltzmann constant. 
A similar scaling variable can be defined in binary mixtures with the additive IPL potentials  \cite{Bernu-Hiwatari-Hansen-Pastore} and even in more general fragile systems that can be mapped onto those of effective IPL systems \cite{Bailey}. 
In Eq. (\ref{effective_volume}), the factor $(\epsilon/T)^{3/\zeta}$ characterizes the degree of the particle overlap, and therefore, we interpret $\phi$ as the effective volume fraction at equilibrium. However, whether the somewhat simplified volume fraction $\varphi$ or the effective volume fraction $\phi$ is used does not matter in a practical sense.  When $n$ is varied at a fixed $T$, $\varphi$ and $\phi$ are essentially the same. On the contrary, when $T$ is varied at a fixed $n$, the physical states are usually characterized not by $\phi$ but by $T$. 

Under an applied flow field, the particle structures and the resultant particle-overlaps are anisotropically modulated according to the given flow symmetry. 
We would like to naively ask whether such a flow-induced anisotropy affects the above scaling. More specifically, we want to know whether the effective volume fraction is changed by varying the shear rate $\dot\gamma$ even at fixed $n$ and $T$. If so, we may further raise a question about how $\phi$ is changed and its impact on the relaxation dynamics. In the following, we provide a detailed explanation of our perspective on these issues and the physical background.

\subsection{Particle configurations under stationary shear flow}

First, let us consider the average particle configurations under shear flow.  
In this study, a simple shear flow with the following mean velocity profile is assumed,  
\begin{eqnarray}
\langle {\mbox{\boldmath$v$}} \rangle={\dot \gamma}y\hat{\mbox{\boldmath$x$}}, \label{shear-flow}
\end{eqnarray}
where the $x$ axis is along the direction of the mean flow, the $y$ axis is along the mean velocity gradient, and $\hat{\mbox{\boldmath$x$}}$ is the unit vector along the $x$ axis. 
For a fixed $T$ condition, $\varphi$ and $\phi$ are essentially the same at $\dot\gamma=0$. 
Therefore, we hereafter reset the reference size and the volume fraction at $\dot\gamma=0$ to be $\sigma$ and $\phi_0=n\pi \sigma^3/6$, respectively. 
Then, we examine how they effectively vary as $\dot\gamma$ changes.

Throughout this study, the average configurations are considered in terms of the two-body pair correlation function. Under the shear flow of Eq. (\ref{shear-flow}), according to the flow symmetry,  $g_s({\mbox{\boldmath$r$}};\phi_0)$ is generally expressed as \cite{Hansen-McdonaldB,Kirkwood-Buff-Green}
\begin{eqnarray} 
g_{s}({\mbox{\boldmath$r$}};\phi_0)=g^{(0)}_s(r;\phi_0)+ {\hat x}{\hat y} g^{(1)}_s(r;\phi_0)+ \cdots, \label{g_of_r}
\end{eqnarray} 
where ${\hat x}=x/r=\sin\psi\cos\theta$ and ${\hat y}=y/r=\sin\psi\sin\theta$, with $\psi$ and $\theta$ representing the polar and azimuthal angles, respectively.   
Here, $g^{(0)}_s(r;\phi_0)$ represents the isotropic part and $ {\hat x}{\hat y}g^{(1)}_s(r;\phi_0)$ is the leading order deviation from $g^{(0)}_s(r;\phi_0)$. 
It is known that $g^{(1)}_s(r;\phi_0)$ is responsible for the non-zero average shear stress, $\Sigma_{xy}$, which is given as \cite{Hansen-McdonaldB,Kirkwood-Buff-Green}
\begin{eqnarray}
\Sigma_{xy}&=& \dfrac{1}{2} n^2  \int d{\mbox{\boldmath$r$}} {\hat x}^2{\hat y}^2 r \dfrac{dU^{\rm (A)}}{dr} g^{(1)}_s(r;\phi_0), 
\label{shear_stress_R}
\end{eqnarray} 
The deviatoric part $g^{(1)}_s(r;\phi_0)$ is approximately given by \cite{Hanley-Rainwater-Hess,Suzuki-Haimovich-Egami,Iwashita-Egami} 
\begin{eqnarray}
g^{(1)}_s(r;\phi_0)  \cong  - c_g \dot\gamma\hat\tau_\alpha  r\dfrac{\partial }{\partial r}g^{(0)}_s(r;\phi_0), \label{configuration1}
\end{eqnarray} 
where $c_g$ is a numerical constant of order unity. 
As shown in the next section, 
this approximate form of $g^{(1)}_s$ with an appropriate value of $c_g$ reproduces the simulation results well. Equation (\ref{configuration1}) can be understood as a consequence of balancing advection and relaxation in the steady state \cite{Hanley-Rainwater-Hess} as follows 
\begin{eqnarray} 
\dot\gamma y \dfrac{\partial }{\partial x} g_s \cong  -\dfrac{1}{\hat\tau_{\alpha}}(g_s-g^{(0)}_s).  \label{eq_motion1}
\end{eqnarray}
By taking the leading order term of $\dot\gamma$, we obtain Eq. (\ref{configuration1}). 
Substituting Eq. (\ref{configuration1}) into Eq. (\ref{g_of_r}) yields 
\begin{eqnarray}
g_s({\mbox{\boldmath$r$}};\phi_0) &\cong& g^{(0)}_s(r;\phi_0)- \lambda  {\hat x}{\hat y} r\dfrac{\partial }{\partial r} g^{(0)}_s(r;\phi_0) \nonumber \\
&\cong& g^{(0)}_s\biggl[\frac{r}{1+ \lambda  {\hat x}{\hat y}};\phi_0\biggr],  \label{g1}
\end{eqnarray}
where  $\lambda=c_g\dot\gamma\hat\tau_\alpha (\lambda\ll 1)$. 
Equation (\ref{g1}) indicates that the average particle configuration under the shear flow is approximately given by distorting the reference configuration $g^{(0)}_s(r;\phi_0)$, i.e., by elongating and compressing by $(1+\lambda/2)$ and $(1-\lambda/2)$ along ${\hat x}{\hat y}=1/2$ and ${\hat x}{\hat y}=-1/2$, respectively. 

\subsection{Reproduction of the sheared configuration by distorting an arbitrary reference}

Next, we note that Eq. (\ref{g1}) can be expressed differently. 
Namely, within the present leading order approximation in $\lambda(\ll 1)$, we can formally rewrite Eq. (\ref{g1}) as  
\begin{eqnarray}
g_{s}({\mbox{\boldmath$r$}};\phi_0)   
\cong  {g}^{(0)}_s\biggl[\dfrac{r}{1 + \lambda ( {\hat x}{\hat y} - b) }; (1-3b\lambda)\phi_0 \biggr],   \label{g4}
\end{eqnarray} 
where we set $b\in [-1/2,1/2]$.  Note that Eq. (\ref{g4}) includes Eq. (\ref{g1}) when $b=0$. 
This formal re-expression of $g_s({\mbox{\boldmath$r$}};\phi_0)$ indicates that  
an identical sheared particle configuration can be obtained by applying a hypothetical distortion 
\begin{eqnarray} 
r\rightarrow \dfrac{r}{1+\lambda(\hat x\hat y -b)}, \label{system_dis}
\end{eqnarray} 
to an isotropic reference configuration described by ${g}_{s}^{(0)}[r; (1-3 \lambda b)\phi_0]$ [different from ${g}_{s}^{(0)}(r;\phi_0)$].  
A more detailed explanation of this reexpression is presented in Appendix A. 

\begin{figure}[th] 
\includegraphics[width=.43\textwidth]{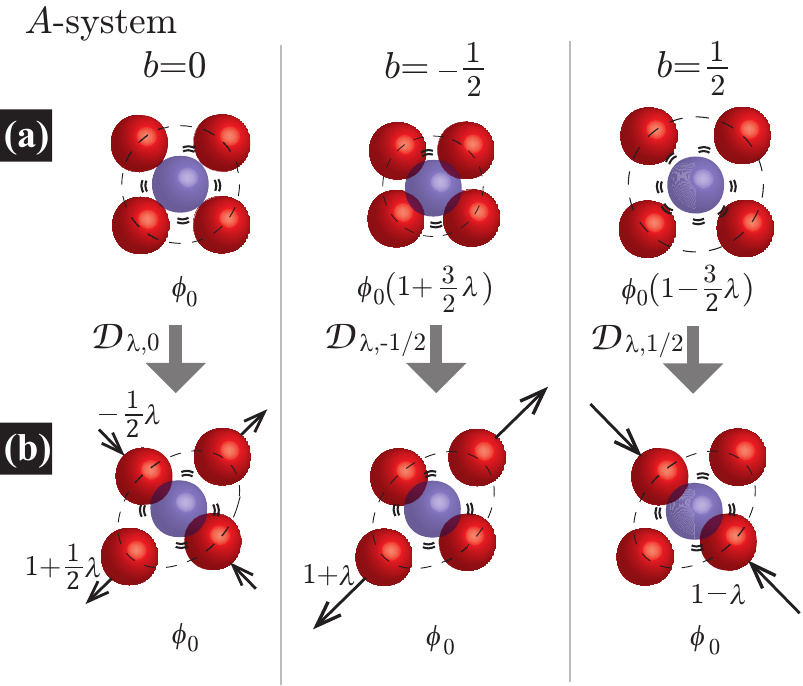}
\caption{(Color online) A schematic showing the reproduction of the sheared structure by anisotropically distorting reference frames. (a) Isotropic particle configuration at three different volume fractions. (b) An almost identical sheared structure is obtained by appropriately distorting the reference frame $\{{\mathcal D}_{\lambda,b}: r\rightarrow {r}/{[1+\lambda(\hat x\hat y -b)]}\}$. For $b=1/2$,  ${\mathcal D}_{\lambda,1/2}$ represents an anisotropic compression, while for 
$b=-1/2$, ${\mathcal D}_{\lambda,-1/2}$ represents an anisotropic expansion. 
When $b=0$, ${\mathcal D}_{\lambda,0}$ represents a pure shear deformation without any volume change. }
\label{OPsystemA}
\end{figure}

By denoting the operation of Eq. (\ref{system_dis}) as ${\mathcal D}_{\lambda,b}$, we may express the present hypothetical operation as   
\begin{eqnarray}
{g}^{(0)}_s[r; (1-3 \lambda b)\phi_0] \xrightarrow{{\mathcal D}_{\lambda,b}} {g}^{(0)}_s\biggl[\dfrac{r}{1 + \lambda ( {\hat x}{\hat y} - b) }; (1-3b\lambda)\phi_0 \biggr]. \nonumber \\  
 \label{operationA}
\end{eqnarray}
These situations are schematically illustrated in Fig. \ref{OPsystemA}. 
Note that, as observers, we generally consider the operation ${\mathcal D}_{\lambda,0}$ to correspond to the actual occurring distortion. 

\subsection{Reproduction of the sheared configuration by distorting constituent particles}

\begin{figure}[bh] 
\includegraphics[width=.4\textwidth]{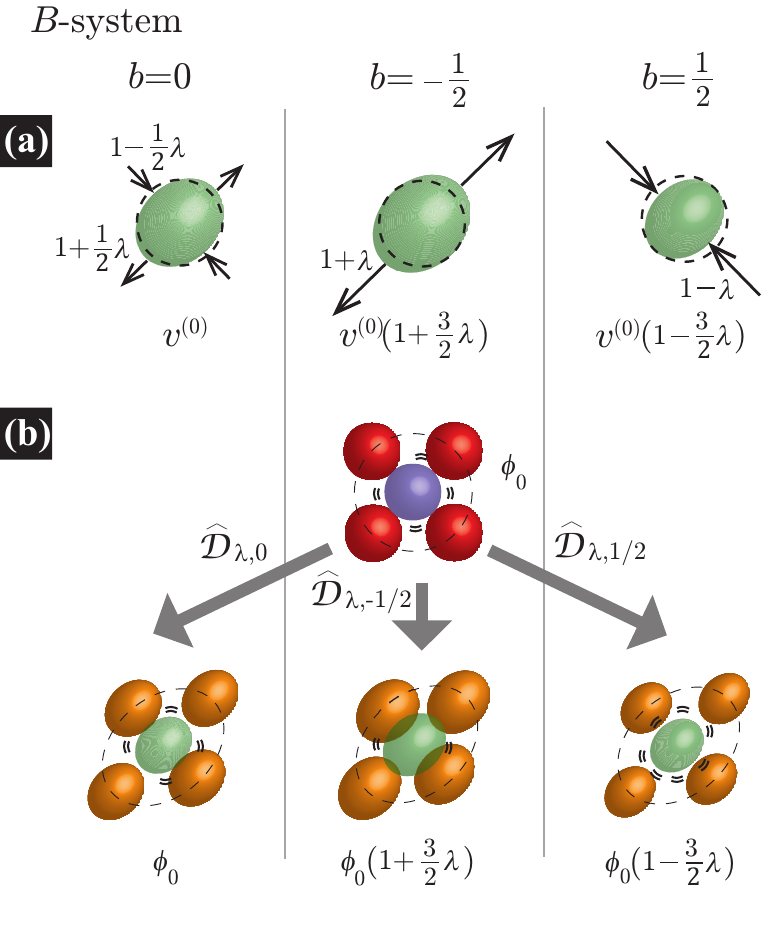}
\caption{(Color online) A schematic showing the reproduction of the sheared particle configuration by anisotropically distorting constituent particles. 
(a)  Anisotropically distorted constituent particles $\{{\hat {\mathcal D}}_{\lambda,b}: \sigma \rightarrow \sigma [1 + \lambda ( {\hat x}{\hat y} - b) ]\}$. (b) Particle configurations obtained by ${\hat {\mathcal D}}_{\lambda,b}$. 
When $b=1/2$, ${\hat {\mathcal D}}_{\lambda,1/2}$ anisotropically reduces the particle size, while when $b=-1/2$, ${\hat {\mathcal D}}_{\lambda,-1/2}$ anisotropically enhances the particle size. 
When $b=0$, ${\hat {\mathcal D}}_{\lambda,0}$ represents particle distortion without changing the particle volume. Changing the particle size, which is controlled by changing $b$, varies the fluctuation and pressure effects. 
These competing fluctuation and pressure effects balance at a fixed number density $n$, resulting in an identical two-body pair correlation function even for different values of $b$.}
\label{OPsystemB}
\end{figure}

In Eq. (\ref{g4}), we consider the {\it system} to be anisotropically distorted while the constituent {\it particles} remain undistorted. 
However, we may interpret Eq. (\ref{g4}) differently: {\it particles} are anisotropically distorted, while the {\it system} remains undistorted. 
That is, with $\sigma$ being the reference size, the particle size is anisotropically modulated as 
\begin{eqnarray}
\sigma \rightarrow \sigma [1 + \lambda ( {\hat x}{\hat y} - b) ],  \label{core}
\end{eqnarray} 
which is assumed to be set by the following interaction potential 
\begin{eqnarray}
{U}^{(B)}({\mbox{\boldmath$r$}})=
\epsilon \biggl\{\dfrac{\sigma[1 + \lambda ( {\hat x}{\hat y} - b) ]}{r}\biggr\}^{\zeta}. 
\label{apotentialB}
\end{eqnarray}  
The particle volume and the volume fraction are given as 
\begin{eqnarray}
\dfrac{1}{24}\int  d\theta d\psi \sin\theta \sigma^3 [1 + \lambda ( {\hat x}{\hat y} - b) ]^3 \cong v^{(0)}(1-3\lambda b), 
\end{eqnarray}
and 
\begin{eqnarray}
\dfrac{Nv^{(0)}}{V}(1-3b\lambda)=\phi_0(1-3b\lambda), 
\end{eqnarray} 
respectively. 
Hereafter, a system where constituent particles interact via $U^{(B)}({\mbox{\boldmath$r$}})$ is referred to as a $B$-system.  
By denoting the operation of Eq. (\ref{core}) as ${\hat {\mathcal D}}_{\lambda,b}$, we may express the present hypothetical operation, Eq. (\ref{core}), as 
\begin{eqnarray}
{g}^{(0)}_s(r;\phi_0) \xrightarrow{{\hat {\mathcal D}}_{\lambda,b}}  {g}^{(0)}_s\biggl[\dfrac{r}{1 + \lambda ( {\hat x}{\hat y} - b) }; (1-3b\lambda)\phi_0 \biggr]. 
\label{operationB}
\end{eqnarray} 
The situations for three different values of $b$ are schematically displayed in Fig. \ref{OPsystemB}.  
In the next section, we demonstrate that the two-body pair correlation function of the $B$-system obtained by MD simulations for various values of $b$ and $\lambda$ nearly exactly reproduces the stationary sheared structure. 

Assuming that the virial theorem holds in the present $B$-system, the compressibility factor $p/nT$ is expressed as 
\begin{eqnarray}
&&\dfrac{p}{nT}-1 \cong \nonumber \\
&&\dfrac{\zeta}{6T}n(1-3\lambda b)\int {\rm d}\Omega' {\rm d}r' U^{(A)}(r')   g^{(0)}_s[r';(1-3b\lambda)\phi_0 ]. \nonumber
\\\label{pressure} 
\end{eqnarray}
A more detailed derivation is presented in Appendix B. 
In Eq. (\ref{pressure}), $p$ formally represents the equilibrium pressure of the $A$-system at a volume fraction of $(1-3\lambda b)\phi_0$ and a number density of $n$. The pressure is smaller for larger $b$ at a fixed number density $n$. 
Furthermore, the pressure component along arbitrary directions does not depend on the direction and, thus, is isotropic. This resultant isotropy reflects the absence of distortion of the system itself. 
Based on the view presented in this subsection, Eq. (\ref{g4}) implies that the configurations of differently modulated particles [by Eq. (\ref{core})] share almost identical two-body correlation functions \cite{comment_example}. This can be naively understood as follows. At a fixed system size, enhancing the particle size (decreasing $b$) reduces fluctuations but increases the pressure. These two competing pressure and fluctuation effects balance, and the resultant particle configuration remains almost unchanged. 
 
\subsection{Tuning the overlap by tuning the potential}

In the $B$-system, the equation of motion of the $i$-th particle's dynamics may be simply given by, 
\begin{eqnarray}
m\dfrac{\rm d^2}{{\rm d} t^2}{\mbox{\boldmath$R$}_i} =  -\dfrac{\partial }{\partial {\mbox{\boldmath$R$}_i}}\sum_{j\ne i} {U}^{(B)}({\mbox{\boldmath$R$}_{ij}}),  \label{eqB}
\end{eqnarray}
where ${\mbox{\boldmath$R$}_{i}}$ is the position of the $i$-th particle and ${\mbox{\boldmath$R$}_{ij}}={\mbox{\boldmath$R$}_{i}}-{\mbox{\boldmath$R$}_{j}}$. Equation (\ref{eqB}) is formally rewritten as 
\begin{eqnarray}
m\dfrac{\rm d^2}{{\rm d} t^2}{\mbox{\boldmath$R$}_i}= -\dfrac{\partial }{\partial {\mbox{\boldmath$R$}_i}} \sum_{j\ne i}{U}^{(A)}(R_{ij}) + {\mbox{\boldmath$F$}_i}, \label{eqA}
\end{eqnarray}
where  $R_{ij}=|{\mbox{\boldmath$R$}_{i}}-{\mbox{\boldmath$R$}_{j}}|$ and 
\begin{eqnarray}
{\mbox{\boldmath$F$}_i} = - \dfrac{\partial }{\partial {\mbox{\boldmath$R$}_i}} \sum_{j\ne i}[{U}^{(B)}({\mbox{\boldmath$R$}_{ij}}) -  {U}^{(A)}(R_{ij})]. \label{Fex}
\end{eqnarray} 
Equations (\ref{eqB}) and (\ref{eqA}) correspond to the views described in Secs. II D and C, respectively.  
That is, regarding the equation of motion, the difference in these views corresponds to the difference in whether ${\mbox{\boldmath$F$}_i}$ is regarded as the {\it intrinsic} or {\it extrinsic} force.   
Note that neither Eqs. (\ref{eqB}) nor (\ref{eqA}) describe the actual particle dynamics under shear flow. Still, these dynamics may reproduce the stationary structural configurations of the sheared system, as shown in the next section.  

In Secs. II B-D, we discussed that the ``average'' shear-flow effects on the particle configurations can be reproduced by applying hypothetical distortions to the {\it system} or the {\it particles} even without shear flow. 
However, in reproducing the steady structure, $g_{s}({\mbox{\boldmath$r$}};\phi_0)\cong g_s^{(0)}\{{r}/{[1+\lambda({\hat x}{\hat y}-b)]};(1-3b\lambda)\phi_0 \}$, there is an arbitrariness in choosing $b$. In the next section, we use MD simulations to demonstrate that only a specific operation with $b=1/2$ can appropriately reproduce the actual relaxation dynamics, removing such arbitrariness.

\section{Numerical results}
In this section, following the arguments presented in the previous section, we perform MD simulations to demonstrate that the structure and the dynamics of the sheared system can be mapped onto a system in which the constituent particles interact via the anisotropically modulated potential $U^{(B)}$ with a specific value of $b(=1/2)$.

\subsection{Simulations of the sheared $A$-system with isotropic potentials}

In our simulations, we employ a binary mixture of large ($L$) and small ($S$) particles interacting via the (soft core) IPL potentials given by \cite{Bernu-Hiwatari-Hansen,Bernu-Hiwatari-Hansen-Pastore,Roux-Barrat-Hansen}
\begin{eqnarray}
U_{\mu\nu}^{(A)}(r)=\epsilon \biggl(\dfrac{\sigma_{\mu\nu}}{r}\biggr)^{12},  \label{potentialAmd}
\end{eqnarray}
where $\mu,\nu= L,S$ and $r$ is the distance between two particles. 
$\sigma_{\mu\nu}=(\sigma_{\mu}+\sigma_{\nu})/2$, where $\sigma_{\mu}$ is conveniently set to the size of the $\mu$ species particle in the reference state; that is, similar to the setting of the particle size in the monodisperse case (Sec. IIA), we also set the reference particle size when $\dot\gamma=0$ to be $\sigma_\mu$ ($\mu=L,S$). Under this setting, the reference particle volume and the volume fraction are given as 
\begin{eqnarray}
 v_\mu^{(0)} = \dfrac{1}{6}\pi \sigma_\mu^3, 
\end{eqnarray}
and 
\begin{eqnarray}
\phi_0=\dfrac{N_{\rm L}v_{\rm L}^{(0)}+N_{\rm S}v_{\rm S}^{(0)}}{V}, \label{volume_fraction}
\end{eqnarray} 
respectively. 
The mass and size ratios are $m_{L}/m_{S}=2$ and $\sigma_{L}/\sigma_{S}=1.2$, respectively. The units for length and time are $\sigma_S$ and $({m_{S}\sigma_{S}^{2}/\epsilon})^{1/2}$, respectively. The total number of particles is  $N=N_{L}+N_{S}=8000$ and $N_L/N_S=1$. The temperature $T$ is measured in units of $\epsilon/k_{\rm B}$. The fixed particle number density and the linear dimension of the system are $N/V =0.8/\sigma_{S}^{3}$ and $L=21.54$, respectively. In this simulation, under simple shear flow, Eq. (\ref{shear-flow}), 
the equations of motion are solved using Lee-Edwards periodic boundary conditions with a Gaussian thermostat \cite{RapaportB}. 

\begin{figure}[hbt] 
\includegraphics[width=.45\textwidth]{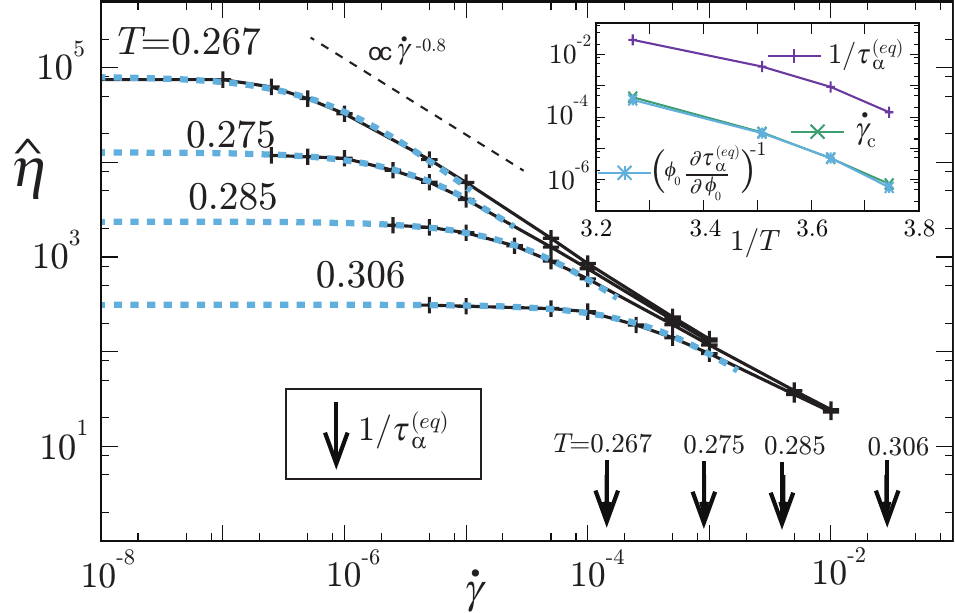}
\caption{(Color online) The main panel shows the steady-state shear viscosity $\hat\eta$ as a function of the shear rate $\dot\gamma$ for several temperatures. The crossover shear rate from Newtonian to non-Newtonian behavior $\dot\gamma_c$ is determined by the fit, $\eta^{(eq)}/(1+\dot\gamma/\dot\gamma_c)$, which is indicated by the dashed lines. 
For $\dot\gamma/\dot\gamma_{c}\gg 1$, $\hat\eta\sim \dot\gamma^{-p}$. The shear-thinning exponent $p$ is less than 1 (at $T=0.267$, $p\cong 0.8$) and seems to show a slight $T$ dependence. 
The arrows represent $1/\tau_\alpha^{(eq)}$, indicating $\dot\gamma_c\tau\alpha^{(eq)}\ll 1$. 
In the inset, we plot $1/\tau_\alpha^{(eq)}$, $\dot\gamma_c$, and the theoretically predicted crossover shear rate $\phi_0(\partial \tau_\alpha^{(eq)}/\partial \phi_0)^{-1}$ (Eq. (\ref{crossover_fragile}) derived below in Sec. IIIC) against $1/T$. 
We find that  $\dot\gamma_c$ quantitatively corresponds with $\phi_0(\partial \tau_\alpha^{(eq)}/\partial \phi_0)^{-1}$. }
\label{geta}
\end{figure}

In Fig. \ref{geta}, we show the $\dot\gamma-\hat\eta$ curves for the present model, which exhibit shear-thinning behavior. 
Crossovers from Newtonian to non-Newtonian flow behavior at $\dot\gamma\tau^{(eq)}>1$ with $\tau^{(eq)}$ being the equilibrium relaxation time are observed in many soft matter systems: $\dot\gamma\tau^{(eq)}>1$ indicates the dominance of advective effects over equilibrium structural relaxation mechanisms (the so-called constitutive instability) in flows. Therefore, similar crossovers might be expected to occur in glass-forming liquids. 
However, as shown in Fig. \ref{geta}, by focusing on the average degree of the shear ``distortion'' at the crossover, we find that shear thinning starts when $\dot\gamma$ is several orders of magnitude smaller than $1/\tau_\alpha^{(eq)}$, which indicates quite a small average structural distortion ($\lambda\ll  10^{-2}$). This is also the case in most experiments \cite{Webb-Dingwell,Kato-Kawamura-Inoue-Chen,Lu-Ravichandran-Johonson} and simulations \cite{Yamamoto-Onuki,Berthier-Barrat,Varnik,Lemaitre,FurukawaS1,Lubchenko,FurukawaS3,FurukawaS4,Trond-Tanaka,Ghosh-Schweizer} of supercooled liquids, where the onset of shear thinning occurs at approximately $\dot\gamma{\tau}_\alpha^{(eq)} \sim 10^{-2} \sim 10^{-3}$.  
This large time-scale separation may exclude the possibility of the usual constitutive instability \cite{FurukawaS1,Lubchenko,FurukawaS3,FurukawaS4}, and, therefore, is an important characteristic of rheological features observed near the crossover from Newtonian to non-Newtonian flow behaviors. In the standard simulation of supercooled liquids, the Lindemann length is typically approximately 0.1 times the particle size; therefore, the amplitude of strain fluctuations at the particle scale is much larger than the average shear distortion $\dot\gamma{\hat \tau}_\alpha$. In such a situation, the shear-flow effects should be much weaker than the thermal fluctuation effects and are regarded as a small perturbation to the structures. However, as shown in Sec. III C, they should have a strong impact on the dynamics.  

\begin{figure}[hbt] 
\includegraphics[width=.48\textwidth]{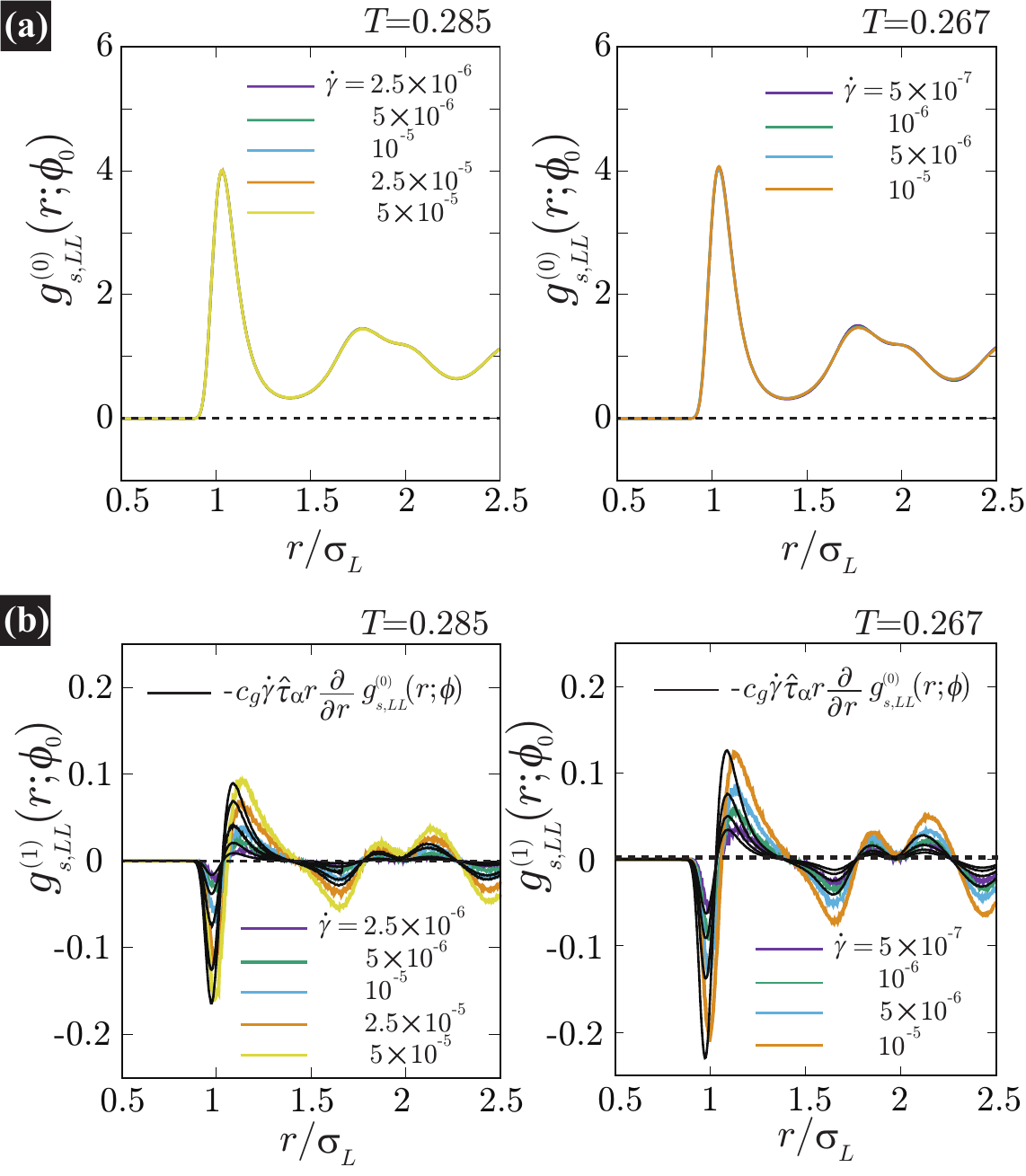}
\caption{ 
(Color online) $g_{s,LL}^{(0)}(r;\phi_0)$ (a) and $g_{s,LL}^{(1)}(r)$ (b) for various shear rates at $T=0.285$ (left) and $0.267$ (right). 
To leading order in $\dot\gamma$, the pair correlation function $g_{s,LL}({\mbox{\boldmath$r$}};\phi_0)$ is expressed as $g_{s,LL}({\mbox{\boldmath$r$}};\phi_0) $=$ g_{s,LL}^{(0)}(r;\phi_0)+ {\hat x}{\hat y} g_{s,LL}^{(1)}(r;\phi_0)$.  
In (a), $g_{s,LL}^{(0)}(r;\phi_0)$ collapses onto a single curve.  
As shown in (b), for particle pairs in the first shell, $g_{s,LL}^{(1)}(r)$ is well approximated by $g_{s,LL}^{(1)}(r)\cong - \lambda  r({\partial }/{\partial r}) g_{s,LL}^{(0)}(r;\phi_0)$, where $\lambda=c_g\dot\gamma\hat\tau_\alpha(\ll 1)$, and we set $c_g=0.65$ and 0.45 for $T=0.285$ and 0.267, respectively.  
Almost the same results are obtained for $g_{s,SS}({\mbox{\boldmath$r$}};\phi_0)$ and $g_{s,SL}({\mbox{\boldmath$r$}};\phi_0)$ as those for $g_{s,LL}({\mbox{\boldmath$r$}};\phi_0)$ using the same value of $c_g$. 
}
\label{structureA}
\end{figure}

Similar to Eq. (\ref{g_of_r}) for the monodisperse case, under the shear flow of Eq. (\ref{shear-flow}), the pair correlation functions $g_{s,\mu\nu}({\mbox{\boldmath$r$}};\phi_0)$ are expressed as \cite{Kirkwood-Buff-Green} 
\begin{eqnarray} 
g_{s,\mu\nu}({\mbox{\boldmath$r$}};\phi_0)=g_{s,\mu\nu}^{(0)}(r;\phi_0)+ {\hat x}{\hat y} g_{s,\mu\nu}^{(1)}(r;\phi_0)+ \cdots, \label{g_of_r_md}
\end{eqnarray} 
where $g_{s,\mu\nu}^{(0)}(r;\phi_0)$ represents the isotropic part and $ {\hat x}{\hat y}g_{s,\mu\nu}^{(1)}(r;\phi_0)$ is the leading order deviation from $g_{s,\mu\nu}^{(0)}(r;\phi_0)$. 
Similar to Eqs. (\ref{configuration1}) and (\ref{g1}) presented in Sec. II B, the deviatoric part $g_{s,\mu\nu}^{(1)}(r)$ is approximately given as \cite{Hanley-Rainwater-Hess,Suzuki-Haimovich-Egami,Iwashita-Egami} 
\begin{eqnarray}
g_{s,\mu\nu}^{(1)}(r;\phi_0)  \cong  - c_g \dot\gamma\hat\tau_\alpha  r\dfrac{\partial }{\partial r}g_{s,\mu\nu}^{(0)}(r;\phi_0), \label{configuration1md}
\end{eqnarray} 
and we obtain 
\begin{eqnarray}
g_{s,\mu\nu}({\mbox{\boldmath$r$}};\phi_0) &\cong& g_{s,\mu\nu}^{(0)}(r;\phi_0)- \lambda  {\hat x}{\hat y} r\dfrac{\partial }{\partial r} g_{s,\mu\nu}^{(0)}(r;\phi_0) \nonumber \\
&\cong& g_{s,\mu\nu}^{(0)}\biggl[\frac{r}{1+ \lambda  {\hat x}{\hat y}};\phi_0\biggr].  \label{g1md}
\end{eqnarray}
Here, $c_g$ is a numerical constant of order unity and 
$\lambda=c_g\dot\gamma\hat\tau_\alpha (\lambda\ll 1)$. As shown in Fig. \ref{structureA}(b), for the present model system, this approximate form of $g_{s,\mu\nu}^{(1)}$, with $c_g$ being a constant of order unity \cite{comment_cg}, nearly reproduces the simulation results. 

Within the present leading order approximation in $\lambda(\ll 1)$, Eq. (\ref{g1md}) can also be expressed as  
\begin{eqnarray}
g_{s,\mu\nu}({\mbox{\boldmath$r$}};\phi_0)  
\cong  {g}_{s,\mu\nu}^{(0)}\biggl[\dfrac{r}{1 + \lambda ( {\hat x}{\hat y} - b) }; (1-3b\lambda)\phi_0 \biggr],   \label{g4md}
\end{eqnarray} 
where $b\in [-1/2,1/2]$. 
Equation (\ref{g4md}) includes Eq. (\ref{g1md}) when $b=0$. 
The meaning of this formal reexpression has been already discussed for the monodisperse case in Sec. II.

\subsection{Simulations of the unsheared $B$-system with anisotropic potentials} 

As discussed in Secs. II C and D, there are two ways to interpret Eq. (\ref{g4md}). 
One is that the {\it system} is anisotropically distorted while the {\it particles} remain undistorted. 
This operation is denoted as ${ {\mathcal D}}_{\lambda,b}: r \rightarrow r/[1 + \lambda ( {\hat x}{\hat y} - b) ]$
The other is that the {\it particles} are anisotropically distorted, while the {\it system} remains undistorted, denoted as ${\hat{\mathcal D}}_{\lambda,b}: \sigma \rightarrow \sigma[1 + \lambda ( {\hat x}{\hat y} - b) ]$: 
the distorted and undistorted terms are simply  interchanged by taking relative views of ${{\mathcal D}}_{\lambda,b}$ and ${\hat{\mathcal D}}_{\lambda,b}$. 

In this subsection, we show that the structure and the relaxation dynamics of the system obtained by the hypothetical operation ${\hat{\mathcal D}}_{\lambda,1/2}$ can reproduce those of the actual sheared system. 
For this purpose, let us consider the following interaction potentials for binary mixtures 
\begin{eqnarray}
{U}^{(B)}_{\mu\nu}({\mbox{\boldmath$r$}}_{ij})=
\epsilon \biggl\{\dfrac{\sigma_{\mu\nu}[1 + \lambda ( {\hat x}_{ij}{\hat y}_{ij} - b) ]}{r_{ij}}\biggr\}^{12}, 
\label{apotentialBmd}
\end{eqnarray}  
with the effective particle size of the $\mu$-species being anisotropically modulated as 
\begin{eqnarray}
\sigma_{\mu} \rightarrow \sigma_\mu [1 + \lambda ( {\hat x}{\hat y} - b) ].  \label{coremd}
\end{eqnarray} 
Here, $\lambda$ and $b$ are the parameters controlling the degree of distortion and the size of the particles, respectively, as in the monodisperse case discussed in Sec. II D.  The particle volume and the volume fraction are given as 
\begin{eqnarray}
\int  d\theta d\psi \sin\theta \sigma_\mu^3 [1 + \lambda ( {\hat x}{\hat y} - b) ]^3 \cong v_\mu^{(0)}(1-3\lambda b), 
\end{eqnarray}
and 
\begin{eqnarray}
\dfrac{N_{\rm L}v_{\rm L}^{(0)}+N_{\rm S}v_{\rm S}^{(0)}}{V}(1-3b\lambda)=\phi_0(1-3b\lambda),  \label{volume_fractionB}
\end{eqnarray} 
respectively. The other settings are the same as those of the $A$-system presented in Sec. III A. We simulate the present model, where constituent particles interact via ${U}^{(B)}_{\mu\nu}({\mbox{\boldmath$r$}}_{ij})$ without shear flow, using velocity Verlet algorithms in the NVE ensemble \cite{RapaportB}. 

Before proceeding, we note the following. 
Since the off-diagonal components of the stress tensor are not symmetric due to the asymmetric form of $U^{(B)}$, the net torque is not exactly zero. However, the particle configurations are distorted so that the resultant local torques are sufficiently suppressed. 
Therefore, there are no strange rotational motions in the $B$-system. 
Furthermore, in both the sheared $A$- and unsheared $B$-systems, as long as $\lambda$ is sufficiently small, the dynamics are almost isotropic.  
For the sheared $A$-system, the deviatoric particle motions, in which the contribution from convective transport by the average shear flow is subtracted, show minimal marked anisotropy at the two-body correlator level \cite{Miyazaki-Yamamoto-Reichman}. 
However, although some anisotropies emerge from longer-time behaviors and are captured in the dynamic heterogeneity or shear bandings \cite{FurukawaS1}, their roles in the rheological properties remain poorly understood.

\subsubsection{Structures}

\begin{figure}[hbt] 
\includegraphics[width=.46\textwidth]{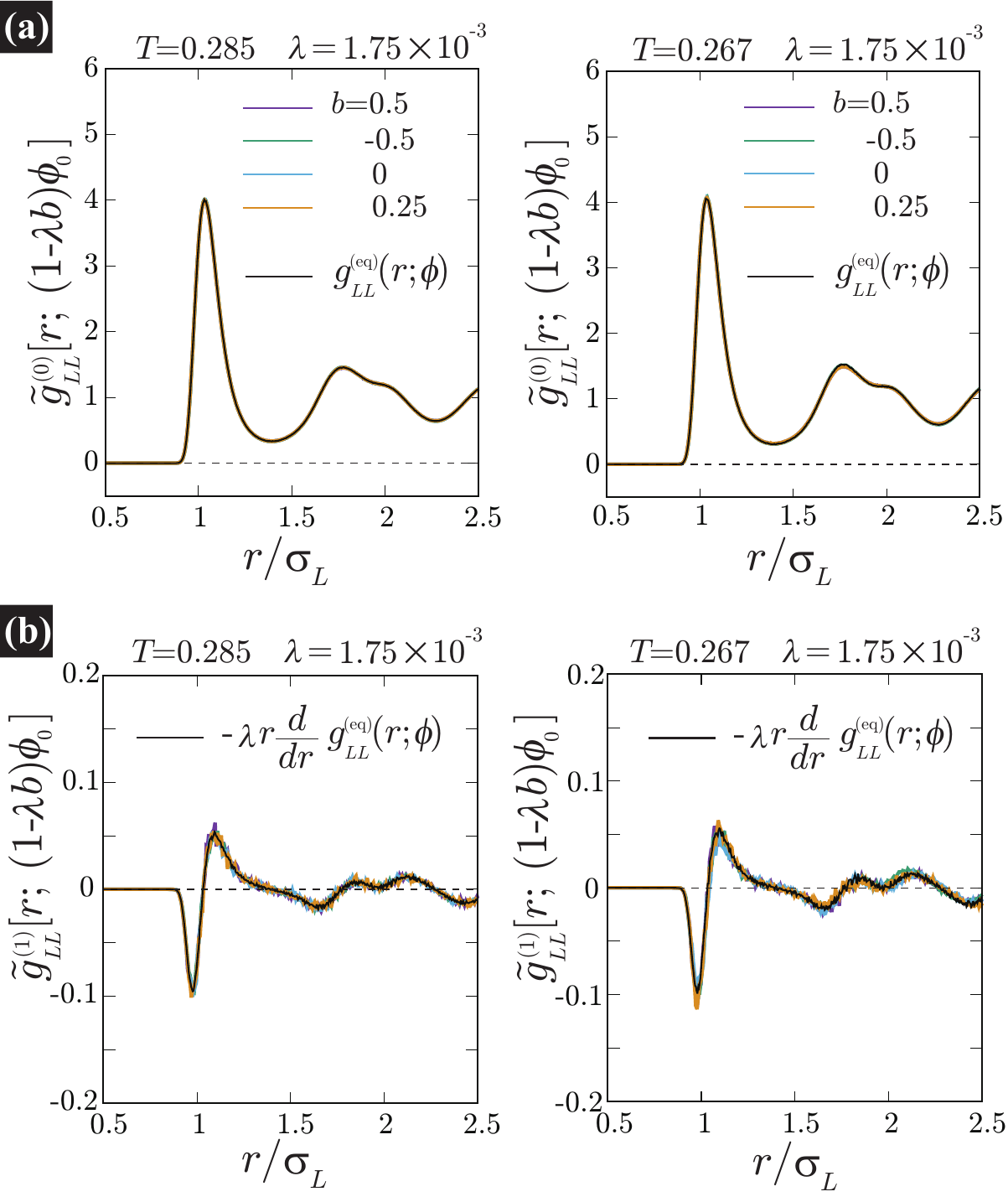}
\caption{ 
(Color online) ${\tilde g}_{LL}^{(0)}[r;(1-b\lambda)\phi_0]$ (a) and ${\tilde g}_{LL}^{(1)}[r;(1-b\lambda)\phi_0]$ (b) for several values of $b$ at $T=0.285$ (left) and 0.267 (right). 
To leading order in $\lambda$, the pair correlation function is expressed as ${\tilde g}_{LL}[{\mbox{\boldmath$r$}};(1-b\lambda)\phi_0]$=$ {\tilde g}_{LL}^{(0)}[r;(1-b\lambda)\phi_0]+ {\hat x}{\hat y} {\tilde g}_{LL}^{(1)}[r;(1-b\lambda)\phi_0]$.  
In (a), ${\tilde g}_{LL}^{(0)}[r;(1-b\lambda)\phi_0]$ collapses onto a single curve and corresponds to ${g}_{LL}^{\rm (eq)}(r;\phi_0)$, which is the equilibrium pair correlation function of the $A$-system.  
As shown in (b),  ${\tilde g}_{LL}^{(1)}[r;(1-b\lambda)\phi_0]$ is well approximated by ${\tilde g}_{LL}^{(1)}[r;(1-b\lambda)\phi_0]\cong - \lambda  r({\partial }/{\partial r}) g_{\mu\nu}^{\rm (eq)}(r;\phi_0)$.  
Almost the same results are obtained for ${\tilde g}_{SS}[{\mbox{\boldmath$r$}};(1-b\lambda)\phi_0]$ and ${\tilde g}_{SL}[{\mbox{\boldmath$r$}};(1-b\lambda)\phi_0]$. 
}
\label{structureB}
\end{figure}

 Hereafter, we denote the two-body correlation function of the $B$-system explicitly as ${\tilde g}_{\mu\nu}[{\mbox{\boldmath$r$}};(1-3b\lambda)\phi_0]$. 
 Considering the symmetry of $U^{(B)}_{\mu\nu}({\mbox{\boldmath$r$}})$, ${\tilde g}_{\mu\nu}[{\mbox{\boldmath$r$}};(1-3b\lambda)\phi_0]$ is expressed as 
 \begin{eqnarray}
&& {\tilde g}_{\mu\nu}[{\mbox{\boldmath$r$}};(1-b\lambda)\phi_0] = \nonumber \\
 &&  {\tilde g}_{\mu\nu}^{(0)}[r;(1-b\lambda)\phi_0] +{\hat x}{\hat y} {\tilde g}_{\mu\nu}^{(1)}[r;(1-b\lambda)\phi_0] +\cdots. 
 \end{eqnarray}
In Fig. \ref{structureB}, we plot ${\tilde g}_{LL}[{\mbox{\boldmath$r$}};(1-b\lambda)\phi_0]$ for various values of $b$ and $\lambda$.  
We find ${\tilde g}_{LL}^{(0)}[r;(1-b\lambda)\phi_0] \cong {g}_{LL}^{\rm (eq)}(r;\phi_0)$ 
and ${\tilde g}_{LL}^{(1)}[r;(1-b\lambda)\phi_0] \cong -\lambda r ({\partial}/{\partial r}) {g}_{LL}^{\rm (eq)}(r;\phi_0)$, where ${g}_{\mu\nu}^{\rm (eq)}(r;\phi_0)$ is the pair correlation function of the $A$-system at equilibrium ($\dot\gamma=0$). 
Note that almost the same results are obtained for ${\tilde g}_{SS}$ and ${\tilde g}_{SL}$ as those obtained for ${\tilde g}_{LL}$. 

As discussed for the monodisperse case in Sec. II, Fig. \ref{structureB} shows that the particle configurations of the $B$-system for different $b(\in [-1/2,1/2])$ have almost identical two-body correlators that are approximately described as  
\begin{eqnarray}
 {\tilde g}_{\mu\nu}[{\mbox{\boldmath$r$}};(1-b\lambda)\phi_0] &\cong&  
  {g}_{\mu\nu}^{\rm (eq)}(r;\phi_0)  - {\hat x}{\hat y}\lambda r \dfrac{\partial}{\partial r} {g}_{\mu\nu}^{\rm (eq)}(r;\phi_0) \nonumber \\ 
  &\cong& {g}_{\mu\nu}^{\rm (eq)}\biggl[ \dfrac{r}{1+\lambda {\hat x}{\hat y}};\phi_0\biggr].  \label{g1mdB}
\end{eqnarray}
For the sheared $A$-system, as long as $\lambda=c_g{\dot\gamma}{\hat \tau}_\alpha\ll 1$, 
$g_{s,\mu\nu}^{(0)}(r;\phi_0) \cong g_{\mu\nu}^{\rm (eq)}(r;\phi_0)$. 
Therefore, from Eqs. (\ref{g1md}) and (\ref{g1mdB}), we deduce 
\begin{eqnarray}
{\tilde g}_{\mu\nu}[{\mbox{\boldmath$r$}};(1-b\lambda)\phi_0] \cong { g}_{s,\mu\nu}({\mbox{\boldmath$r$}};\phi_0),  
\end{eqnarray}
from which we may conclude that the unsheared $B$-system can approximately reproduce the average particle configurations of the sheared $A$-system.

\subsubsection{Dynamics}

\begin{figure}[bt] 
\includegraphics[width=.44\textwidth]{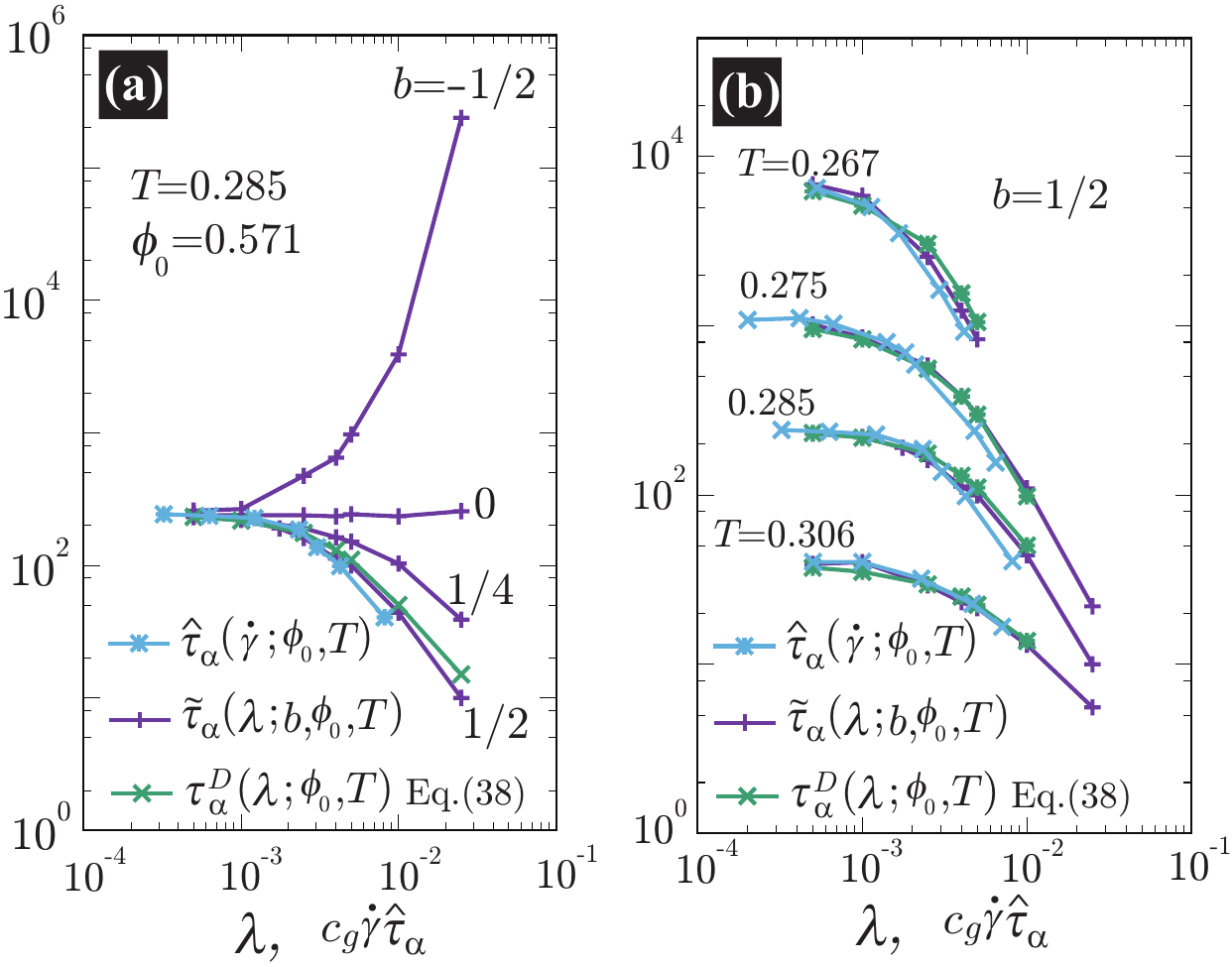}
\caption{ 
(Color online) 
(a) The structural relaxation time of the $B$-system ${\tilde \tau}_\alpha(\lambda;b,\phi_0,T)$ is plotted against $\lambda$ for several values of $b$.  Even in systems with identical correlation functions as shown in Fig. \ref{structureB}, a slight difference in the volume fraction results in a notable difference in ${\tilde \tau}_\alpha(\lambda;b,\phi_0,T)$. 
When $b=-1/2$, ${\tilde \tau}_\alpha(\lambda;b,\phi_0,T)$ increases with an increase in $\lambda$ due to the increase in the volume fraction as $\phi_0(1+3\lambda/2)$. In contrast, for $b=1/4$ and $1/2$, the opposite result occurs. When $b=0$, the relaxation time remains unchanged, suggesting that a small anisotropy ($\lambda\lesssim 10^{-2}$) without changing the volume fraction does not affect structural relaxation. 
The relaxation time of the sheared $A$-system ${\hat\tau}_\alpha(\dot\gamma;\phi_0,T)$ is also plotted against $\lambda(=c_g\dot\gamma {\hat\tau}_\alpha)$.   
We find that ${\hat\tau}_\alpha(\dot\gamma;\phi_0,T)$ and ${\tilde\tau}_\alpha(\lambda;b=1/2,\phi_0,T)$ nearly coincide with each other.  Furthermore, we find that the Doolittle equation, Eq. (\ref{DoolittleEq}), $\tau^{D}_{\alpha}(\phi_s,T)= \tau_0^{D}\exp[{\Gamma}\phi_s/{(\phi_c-\phi_{s}})]$ with $\phi_{s}=\phi_0(1-3\lambda/2)$, quantitatively agrees with ${\hat\tau}_\alpha(\dot\gamma;\phi_0,T)$ and ${\tilde\tau}_\alpha(\lambda;b=1/2,\phi_0,T)$ at the same $\lambda(\lesssim 10^{-2})$. Here, $\phi_c$, $\tau_0^{D}$, and $\Gamma$ generally depend on $T$. 
These agreements indicate that (i) shear thinning can be attributed to the shear-flow induced reduction of the volume fraction, and (ii) the relaxation dynamics can be mapped onto the equilibrium dynamics with the corresponding reduced volume fraction. (b) The agreement among $\tau^D_{\alpha}(\phi_s,T)$, ${\hat\tau}_\alpha(\dot\gamma;\phi_0,T)$, and ${\tilde\tau}_\alpha(\lambda;b=1/2,\phi_0,T)$ for several temperatures in supercooled states. 
}
\label{relaxation_time}
\end{figure}

In Fig. \ref{relaxation_time}, the structural relaxation time of the $B$-system, which hereafter is denoted as  ${\tilde \tau}_\alpha(\lambda;b,\phi_0,T)$, is plotted against $\lambda$ for several values of $b$. 
In the present study, the structural relaxation time is defined as the relaxation time of the shear-stress autocorrelation function. For more details, please refer to Appendix C. 
At the same $\lambda$, despite almost the same particle configurations for different $b$ (shown in Fig. \ref{structureB}), the behaviors of the structural relaxation times are quite different. 
When $b<0$, ${\tilde \tau}_\alpha(\lambda;b,\phi_0,T)$ increases with an increase in $\lambda$, reflecting the increase of the volume fraction (and the resultant pressure). 
However, when $b>0$, the opposite result occurs. 
Remarkably, when $b=0$, the relaxation time remains unchanged, which suggests that a  small anisotropy ($\lambda\ll 1$) without any volume changes does not affect the structural relaxation. 
In Fig. \ref{relaxation_time}, we also show the relaxation time of the sheared $A$-system ${\hat\tau}_\alpha(\dot\gamma;\phi_0,T)$ against the degree of the average shear distortion $\lambda=c_g\dot\gamma {\hat\tau}_\alpha$. 
We find that ${\hat\tau}_\alpha(\dot\gamma;\phi_0,T)$ and ${\tilde\tau}_\alpha(\lambda;b=1/2,\phi_0,T)$ nearly coincide with each other at various temperatures. 

The observed good agreement between ${\hat\tau}_\alpha(\dot\gamma;\phi_0,T)$ and ${\tilde\tau}_\alpha(\lambda;b=1/2,\phi_0,T)$ suggests that such a shear-flow effect can be identified with the effect of the anisotropic modulation of the particle sizes (in terms of ${\hat {\mathcal D}}_{\lambda,b}$) for $b=1/2$. 
Under ${\hat {\mathcal D}}_{\lambda,b}$, as schematically shown in Fig. \ref{OPsystemB}, when $b=1/2$, the particles are anisotropically modulated as 
\begin{eqnarray}
\sigma_{\mu} \rightarrow \sigma_\mu [1 + \lambda ( {\hat x}{\hat y} - 1/2) ],  \label{coremd}
\end{eqnarray}
for which the size along ${\hat x}{\hat y} = 1/2$ remains unchanged.

\begin{figure}[thb] 
\includegraphics[width=.45\textwidth]{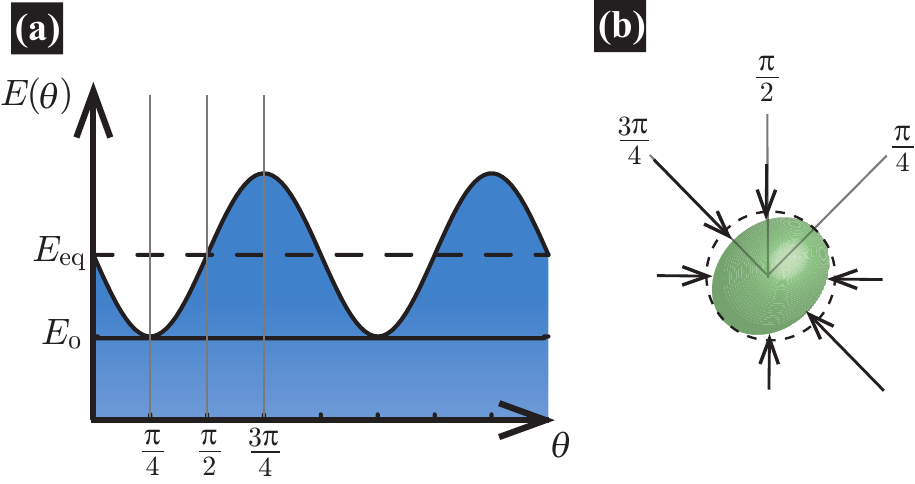}
\caption{ (Color online) 
(a)  A schematic of the shift of the potential energy reference. For the sheared nonequilibrium system, the average interparticle potential energy $E(\theta)$ that a particle experiences is minimal ($E_0$) along the extension axis ($\theta=\pi/4$). Here, $E_{\rm eq}$ is the average value at equilibrium. In the nonequilibrium sheared system, our simulations indicate that 
$E(\theta)-E_0$ is considered to be due to extra overlaps due to the shear flow. (b) A schematic of the anisotropically modulated effective particle size.  The dashed line represents the size at equilibrium.}
\label{schematic_potential}
\end{figure}

We may further interpret Eq. (\ref{coremd}) as follows. 
For the sheared system with varying $\dot\gamma$($\ne 0$) at a fixed $T$, the interparticle potential energy becomes anisotropic: along the extension axis (${\hat x}{\hat y} = 1/2$ in the present case) of the external flow field, dilution occurs, while along the compression axis (${\hat x}{\hat y} = -1/2$), densification occurs. 
Along the direction of ${\hat x}{\hat y}=0$, the particle configurations are not changed from those at equilibrium, whereby as ``observers'', we conventionally set this direction as the reference and consider that the volume fraction to be unchanged with varying $\dot\gamma$.  
However, our simulation results suggest that this conventional setting is not true for ``particles''. Under the external flow field, Eq. (\ref{shear-flow}), as schematically shown in Fig. \ref{schematic_potential}, the average interparticle potential becomes {\it minimal} along the extension axis. 
Then, setting this direction to be the reference, in other directions, the potential energy is ``lifted'' up due to the shear flow, making extra particle overlaps in addition to overlaps due to thermal fluctuations. By subtracting the extra overlap regions, Eq. (\ref{coremd}) with $b=1/2$ describes the effective particle size of the actual sheared system, and the effective volume fraction is given by $\phi_0(1-d\lambda/2)$ at a fixed $T$. Here, $\lambda=c_g\dot\gamma{\hat \tau}_\alpha$ is determined by the distortion of the two-body pair correlation function. 

This shift of the potential energy reference under the shear flow does not alter the observables, such as the average energy, pressure, and shear stress. 
Although our simulations certainly support the present speculation regarding the reduction of the volume fraction induced by the shear flow, the detailed investigation based on first principles are required to provide further evidence.

\subsubsection{Doolittle equation: Mapping onto the equilibrium system with the reduced volume fraction}

Note again that for $\lambda \ll 1$, anisotropy is hardly noticeable in the dynamics at the two-body correlator level \cite{Miyazaki-Yamamoto-Reichman}. We expect that the shear-flow effect is incorporated only through the reduction of the volume fraction and, thus, that the dynamics of the sheared system can be mapped onto the dynamics of the equilibrium system. 

\begin{figure}[hbt] 
\includegraphics[width=.48\textwidth]{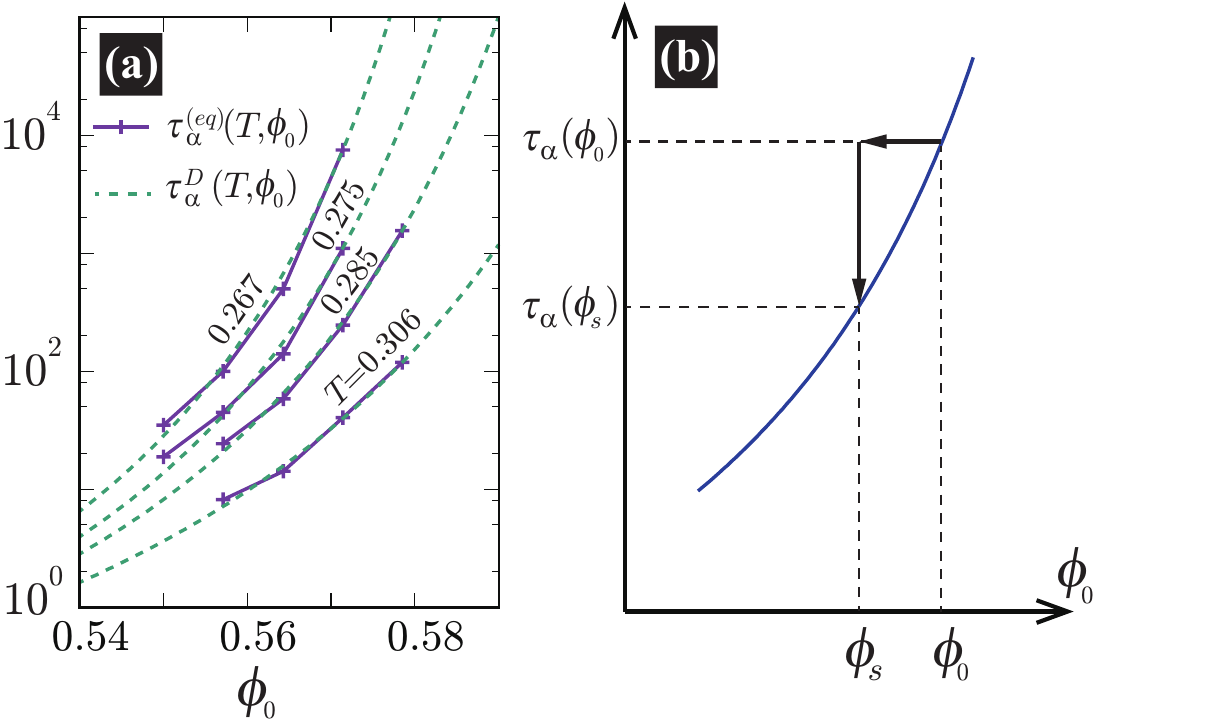}
\caption{ 
(Color online) 
(a) The structural relaxation time of the equilibrium $A$-system $\tau_\alpha^{(eq)}(\phi_0,T)$ for several $T$, which can be fitted to the Doolittle equation (in terms of $\phi_0$) $\tau_\alpha^D(\phi_0,T)=\tau_0^D\exp[\Gamma \phi_0/(\phi_c-\phi_0)]$, represented as the dashed curves. 
(b) A schematic for the acceleration of structural relaxation caused by the shear-induced reduction of the volume fraction. 
As discussed in the main text, due to a small anisotropy at the two-body correlator level, we expect that the dynamics of the sheared system can be mapped onto the equilibrium dynamics by incorporating the shear-flow effect only by reducing the volume fraction: $\phi_0\rightarrow \phi_s=(1-3c_g\dot\gamma \hat\tau_\alpha/2)\phi_0$.  
Close to the glass transition temperature, the volume-fraction dependence of $\tau_\alpha$ becomes much steeper; thus, even a very small decrease in the volume fraction significantly accelerates the relaxation dynamics. 
}
\label{Doolittle}
\end{figure}

In Fig. \ref{relaxation_time}, we also plot the Doolittle equation  \cite{Doolittle1,Doolittle2} 
\begin{eqnarray}
\tau^{D}_{\alpha}(\lambda;\phi_0,T) =\tau^{D}_{\alpha}(\phi_s;T)= \tau_0^{D}\exp\biggl(\dfrac{\Gamma\phi_s}{\phi_c-\phi_{s}}\biggr) \label{DoolittleEq}
\end{eqnarray}
with the reduced volume fraction 
\begin{eqnarray}
\phi_{s}=\phi_0\biggl(1-\dfrac{3}{2}\lambda\biggr)=\phi_0\biggl(1-\dfrac{3}{2} c_g \dot\gamma {\hat \tau}_\alpha\biggr). \label{shift}
\end{eqnarray}
Here, the parameters of $\Gamma$, $\phi_c$, and $\tau_0^{D}$ generally depend on the temperature and are separately determined at equilibrium using MD simulations.  As shown in Fig. \ref{Doolittle}(a), the Doolittle equation approximates the volume-fraction dependence of the structural relaxation time at equilibrium well. 
In Fig. \ref{relaxation_time}, we find that $\tau^D_{\alpha}(\phi_s;T)$ with Eq. (\ref{shift}) quantitatively reproduces the effect of ${\hat\tau}_\alpha(\dot\gamma;\phi_0,T)$ and ${\tilde\tau}_\alpha(\lambda;b=1/2,\phi_0,T)$ with the same $\lambda$, further supporting our perspective. 

Because of a very steep volume-fraction dependence of the relaxation time as described in Eq. (\ref{DoolittleEq}) for supercooled states, even an infinitesimal reduction of the volume fraction causes significant acceleration of the dynamics. 

\subsection{Nonlinear constitutive equations}

As discussed in Sec. IIB, based on the results shown in Fig. \ref{relaxation_time}, we suppose that the relaxation time under shear flow $\hat\tau_\alpha$ is mapped onto the equilibrium $\tau_\alpha$ as  
\begin{eqnarray}
\hat\tau_\alpha(\dot\gamma;\phi_0,T) &=& \tau_\alpha^{\rm (eq)}(\lambda;\phi_0,T) \nonumber \\
&=& \tau_\alpha^{\rm (eq)}\biggl[\phi_0\biggl(1-\dfrac{3}{2}c_g\dot\gamma {\hat \tau}_\alpha\biggr),T\biggr], \label{mapping_fragile}
\end{eqnarray}
where $\phi_0=n(v_L^{(0)}+v_S^{(0)})/2$ from Eq. (\ref{volume_fraction}), and the shear-flow effect is taken into account through the reduced volume fraction $\phi_{s}$. 
This equation is essentially nonlinear in ${\hat \tau}_\alpha$. 
If we know the functional form of $\tau_\alpha^{\rm (eq)}$ (for example, the Doolittle equation), we can solve Eq. (\ref{mapping_fragile}) in terms of ${\hat \tau}_\alpha$. Equation (\ref{mapping_fragile}) can be regarded as the nonlinear constitutive equation and describes the rheological curves of the present model well. 
Since the volume fraction dependence of the shear modulus $G$ is much weaker than that of $\tau_\alpha$, the viscosity may be taken to be 
\begin{eqnarray} 
{\hat \eta}(\dot\gamma;\phi_0,T) \cong G{\hat \tau}_\alpha\cong G(\phi_0,T){\tau}_\alpha^{\rm (eq)}(\phi_{s},T)
\end{eqnarray} 
for $(\phi_0-\phi_s)/\phi_0\ll 1$. Note that the $T$-dependence of $G$ is also much weaker than that of $\tau_\alpha^{\rm (eq)}$.

For $\dot\gamma{\hat\tau_\alpha}\ll 1$, by expanding Eq. (\ref{mapping_fragile}) 
in $\dot\gamma{\hat\tau_\alpha}$, we obtain 
\begin{eqnarray}
\hat\tau_\alpha(\dot\gamma;\phi_0,T) \cong \dfrac{\tau_\alpha^{\rm (eq)}(\phi_0,T)}{1+a \dot\gamma \phi_0 \dfrac{\partial \tau_\alpha^{\rm (eq)}}{\partial \phi_0}}, \label{leading_fragile}
\end{eqnarray}
where $a=3c_g/2$ ($\cong 1$ in the present system).  Therefore, the crossover shear rate from Newtonian to non-Newtonian behavior $\dot\gamma_{\rm c}$ is given by 
\begin{eqnarray}
\dot\gamma_{\rm c} \cong \biggl( \phi_0 \dfrac{\partial \tau_\alpha^{\rm (eq)}}{\partial  \phi_0}\biggr)^{-1}\label{crossover_fragile}. 
\end{eqnarray} 
This crossover shear rate can be much smaller than $1/\tau_\alpha^{\rm (eq)}$ ($\dot\gamma_c\tau_\alpha^{\rm (eq)}\ll 1$) near the glass transition point, indicating that the usual constitutive instability does not trigger the onset of the shear-thinning.
The reference volume fraction $\phi_0$ and the number density $n$ are linearly related to each other ($\phi_0\propto n$), and therefore, Eq. (\ref{crossover_fragile}) is rewritten as 
\begin{eqnarray}
\dot\gamma_{\rm c} \cong \biggl( n \dfrac{\partial \tau_\alpha^{\rm (eq)}}{\partial  n}\biggr)^{-1}\label{crossover_fragile2}. 
\end{eqnarray} 
Equation (\ref{crossover_fragile}) is expressed only in terms of experimental observables, and should thus be useful in the process design of glassy materials. 

Similar predictions for rheological behaviors were obtained in Ref. \cite{FurukawaS3} by more heuristic arguments; here, we rationalize the possible mechanism of the shear-induced reduction of the volume fraction. 
The constitutive equation (\ref{mapping_fragile}) can be quantitatively approximated by the Doolittle equation (\ref{DoolittleEq}) with the reduced volume fraction $\phi_s$ in Eq. (\ref{shift}) or equivalently with the enhanced ``free volume'' $\propto (\phi_c-\phi_s)$. 
Although our perspective on the enhancement of the free volume under shear flow is not exactly the same as the conventional perspective \cite{Spaepen}, we expect that the present study will provide new insights into the physical substance of the free volume.

\section{Concluding remarks}

In this study, we have discussed how the particle size and volume fraction of fragile liquids under shear flow are determined.  
Based on this determination, we derive a nonlinear rheological constitutive equation, which quantitatively describes the shear thinning behavior of fragile supercooled liquids. 

In a shear flow with a shear rate $\dot\gamma$, the particle structures relax with a time scale of ${\hat \tau}_\alpha$, resulting in an average distortion on the order of $\dot\gamma{\hat \tau}_\alpha$.  
The extent to which neighboring particles can overlap is involved in determining the volume fraction; such particle overlap should be controlled by the degree of shear distortion ($\propto \dot\gamma{\hat \tau}_\alpha$) in addition to the strength of thermal fluctuations ($\propto T$).  
Under the simple shear flow of Eq. (\ref{shear-flow}), dilution and densification occur along the extension and compression axes, respectively, with the same magnitude, resulting in the absence of a system volume change. 
In this situation, the overall {\it number density} $n$, which is uniquely determined, is invariant. 
However, unlike the number density, the {\it volume fraction} $\phi$ decreases as $\dot\gamma$ increases. In a system where constituent particles interact via simple short-range repulsive potentials, the average interparticle interactions become minimal along the extension axis (${\hat x}{\hat y}=1/2$). 
By setting the reference direction measuring the potential energy to this extension axis, the potential energy can be considered to be ``lifted'' up in other directions. 
By identifying this lift effect with the shear flow effect inducing extra particle overlaps (in addition to the overlaps due to thermal fluctuations), we obtain Eq. (\ref{coremd}). 
Note that if the reference direction is set to ${\hat x}{\hat y}=0$, which we as {\it observers} usually consider, the particle volume does not vary with the degrees of the shear distortion $\lambda$. However, this may not be true for {\it particles}. Please also refer to the discussion presented at the end of Sec. IIB3. 

Because significant anisotropy is hardly observed in the dynamics at the two-body correlator level \cite{Miyazaki-Yamamoto-Reichman}, we may consider the shear-flow effect to be incorporated primarily by a slight reduction of the volume fraction. Therefore, assuming that the sheared dynamics can be mapped onto the equilibrium dynamics, we obtained the nonlinear constitutive equation, i.e., Eq. (\ref{mapping_fragile}). As clearly shown in Fig. \ref{relaxation_time}, this constitutive equation quantitatively describes the shear-induced acceleration in the relaxation dynamics of fragile supercooled liquids.

Finally, we note the following: 
 
(i) Our preliminary simulations for another standard fragile model liquid, the Kob-Andersen model \cite{Kob-Andersen}, also reproduce almost the same results as those obtained in the present paper.  
Although the model employed in this study assumes pairwise IPL potentials, the interactions of other systems are generally more complicated. 
However, it has been established that the dynamics of a wide class of fragile glass-formers can be reproduced by the corresponding IPL systems \cite{Bailey,DyreR}, for which the effective particle size and volume fraction are simply defined. 

(ii) In Refs. \cite{Furukawa-Tanaka1,Furukawa-Tanaka2}, shear thinning in glassy liquids was discussed in the context of the shear-induced {\it density inhomogeneity}, inspired by the shear-induced phase separation established in polymeric systems \cite{Onuki,Milner,Helfand-Fredrickson}. 
The critical shear rate for the onset of the inhomogeneous flow is given as ${\dot \gamma}_{cr}=(\partial \eta^{\rm (eq)}/\partial p)_T^{-1}$, where $p$ is the pressure and is identified with the shear rate describing the crossover from Newtonian to non-Newtonian behavior. 
This prediction seems to agree with the experimental results for the supercooled melts of Zr-based bulk metallic glass-formers \cite{Lu-Ravichandran-Johonson}. 
However, as was pointed out in Ref. \cite{Lu-Ravichandran-Johonson}, the flow is still homogeneous when $\dot\gamma\sim \dot\gamma_{cr}$, and inhomogeneous flow occurs when $\dot\gamma\gg \dot\gamma_{cr}$, which contradicts the results of Refs. \cite{Furukawa-Tanaka1,Furukawa-Tanaka2}. 
Note that $(\partial \eta^{\rm (eq)}/\partial p)_T\cong G(\partial \tau_\alpha^{\rm (eq)}/\partial p)_T=G(\partial n/\partial p)(\partial \tau_\alpha^{\rm (eq)}/\partial n)=GK_T n(\partial \tau_{\alpha}^{\rm (eq)}/\partial n)$ \cite{FurukawaS3}. 
Here, $K_T$ is the isothermal compressibility and $GK_T\cong 0.3$, which was estimated from experimental results \cite{Lu-Ravichandran-Johonson,Demetriou-Johnson}. Therefore, $\dot\gamma_c$ and $\dot\gamma_{cr}$ are comparable to each other since $\dot\gamma_{c}=0.3\dot\gamma_{cr}$, and the agreement between the predicted $\dot\gamma_{cr}$ and the experimental results may instead indicate the validity of the present mechanism of the shear-induced reduction of the volume fraction, which is not related to the density inhomogeneity. 
We will discuss which mechanism is selected under actual experimental situations in more detail elsewhere. 

(iii) In this study, the considered systems are in (supercooled) liquid states, where thermal fluctuations exert important effects. However, thermal effects are irrelevant in amorphous solid  states. In amorphous states, close links between the shear distortion of microscopic configurations and nonlinear rheological properties have been intensively studied in Refs. \cite{Zaccone1,Zaccone2}. At this stage, it is unclear how our approach for liquid states can be related to amorphous rheology. 

(iv) In a hard core system, particle overlaps never occur. 
However, in this case, what becomes anisotropic is the collision frequency: namely, the collision frequency is enhanced along the compression axis while it is reduced along the extension axis. If we set the extension axis as the reference direction for measuring the collision frequencies, such anisotropies in collisions may be analogous to anisotropic particle overlaps in a soft core system. 

We will examine points (i)-(iv) in future work.

\thanks 

This work was supported by KAKENHI (Grant No. 26103507, No. 25000002, and No. 20508139) and the JSPS Core-to-Core Program ``International research network for nonequilibrium dynamics of soft matter''.

\appendix

\section{}
We formally rewrite Eq. (\ref{g1}) as 
\begin{eqnarray} 
&& g_s({\mbox{\boldmath$r$}};\phi_0) \cong g^{(0)}(r;\phi_0)- \lambda  {\hat x}{\hat y} r\dfrac{\partial }{\partial r} g^{(0)}(r;\phi_0) \nonumber \\ 
&& =\biggl( 1 -  \lambda br\dfrac{\partial}{\partial r}\biggr){g}^{(0)}(r;\phi_0) - \lambda ( {\hat x}{\hat y} -  b )  r\dfrac{\partial}{\partial r}  g^{(0)}(r;\phi_0),  \nonumber\\  \label{Ap_g2} 
\end{eqnarray} 
where we set $b\in [-1/2,1/2]$. 
Actually, we can set $b$ to an arbitrary value, but here we make its range correspond to that of ${\hat x}{\hat y}$ (see also the sentences below Eq. (\ref{Ap_g3})). 
We further rewrite Eq. (\ref{Ap_g2}) by using the following procedure. 
(i) The first term  of Eq. (\ref{Ap_g2}) can be approximated as 
\begin{eqnarray} 
\biggl( 1 -  \lambda br\dfrac{\partial}{\partial r}\biggr){g}^{(0)}(r;\phi_0) &\cong& {g}_{\mu\nu}^{(0)}(\frac{r}{1+\lambda b};\phi_0), \nonumber \\
 &\cong&  {g}^{(0)}[r; (1-3 \lambda b)\phi_0],     \label{Ap_g3} 
\end{eqnarray} 
where, in the first line, according to Eq. (\ref{g1}), ${g}^{(0)}[{r}/{(1+\lambda b)};\phi_0]$ corresponds to $g_s({\mbox{\boldmath$r$}};\phi_0)$ along the direction of ${\hat x}{\hat y} = b$  $(-1/2 \le{\hat x}{\hat y}\le 1/2)$. Then, noting that the average structure viewed in the scaled frame $r\rightarrow r/(1+b\lambda)$ is nearly identical to that at the volume fraction $(1+\lambda b)^{-3} \phi\cong (1-3\lambda b)\phi_0$ in the original frame, we obtain the second line of Eq. (\ref{Ap_g3}). 
(ii) To leading order in $\lambda$, in the second term on the second line of Eq. (\ref{Ap_g2}), $g^{(0)}(r;\phi_0)$ can be replaced by $g^{(0)}[r;(1-3\lambda b)\phi_0]$.  
With these two conditions (i) and (ii), an approximate expression of $g_{s}({\mbox{\boldmath$r$}};\phi_0)$ that is different from Eq. (\ref{g1}) is   
\begin{eqnarray}
g_{s}({\mbox{\boldmath$r$}};\phi_0)  &\cong&  \biggl[ 1 -  \lambda( {\hat x}{\hat y} -  b) r\dfrac{\partial}{\partial r}\biggr] {g}^{(0)}[r; (1-3 \lambda b)\phi_0]   \nonumber\\  
&\cong & {g}^{(0)}\biggl[\dfrac{r}{1 + \lambda ( {\hat x}{\hat y} - b) }; (1-3b\lambda)\phi_0 \biggr].    \label{Ap_g4}
\end{eqnarray}

\section{}

Here, we derive Eq. (\ref{pressure}). 
From the virial theorem, the pressure $p$ of the $B$-system is expressed as 
\begin{eqnarray}
p-nT
&=& -\dfrac{1}{6V}\sum_i^{N}\sum_{j \ne i} \langle {\mbox{\boldmath$R$}}_{ij}\cdot \dfrac{\partial}{\partial {\mbox{\boldmath$R$}}_{ij}}U^{(B)}\rangle  \nonumber \\
&=& \dfrac{\zeta}{6V}\sum_i^{N}\sum_{j \ne i} \langle {U}^{(B)}({\mbox{\boldmath$R$}}_{ij})\rangle \nonumber \\
&=& \dfrac{\zeta n}{6} \sum_{j \ne i} \langle {U}^{(B)}({\mbox{\boldmath$R$}}_{ij})\rangle,  \label{pressure_virial} 
\end{eqnarray}  
where $nT$ represents the ideal gas term, ${\mbox{\boldmath$R$}}_{i}$ is the position of the $i$-th particle, and $ {\mbox{\boldmath$R$}}_{ij} ={\mbox{\boldmath$R$}}_{i}-{\mbox{\boldmath$R$}}_{j}$. 
For a particle located at the origin, the number of particles in an infinitesimal volume element $ {\rm d}{\mbox{\boldmath$r$}}$ centered at the position ${\mbox{\boldmath$r$}}$ is ${\rm d}{\mbox{\boldmath$r$}}[n/(1+\lambda{\hat x}{\hat y} )^3]\times g_{s}({\mbox{\boldmath$r$}};\phi_0)\cong {\rm d}{\mbox{\boldmath$r$}}n(1-3\lambda{\hat x}{\hat y} ) g_{s}({\mbox{\boldmath$r$}};\phi_0)$. 
Therefore, 
\begin{eqnarray}
&&\sum_{j \ne i} \langle {U}^{(B)}({\mbox{\boldmath$R$}}_{ij}) \rangle \nonumber \\
&\cong& \int {\rm d}{\mbox{\boldmath$r$}} \epsilon \biggl\{\dfrac{\sigma[1 + \lambda ( {\hat x}{\hat y} - b) ]}{r}\biggr\}^{\zeta } n(1-3\lambda{\hat x}{\hat y}) g_{s}({\mbox{\boldmath$r$}};\phi_0)
 \nonumber \\
&\cong& \int {\rm d}{\mbox{\boldmath$r$}} \epsilon \biggl\{\dfrac{\sigma[1 + \lambda ( {\hat x}{\hat y} - b) ]}{r}\biggr\}^{\zeta} n(1-3\lambda{\hat x}{\hat y}) \nonumber \\ 
&&\times g^{(0)}\biggl[\dfrac{r}{1+\lambda({\hat x}{\hat y}-b)};(1-3b\lambda)\phi_0 \biggr], \label{pressure_virial2} 
\end{eqnarray}
where Eqs. (\ref{g4}) and (\ref{apotentialB}) have been used. 
Then, by replacing $r/[1+\lambda({\hat x}{\hat y}-b)]$ by $r'$, we obtain 
\begin{eqnarray}
&&\sum_{j \ne i} \langle {U}^{(B)}({\mbox{\boldmath$R$}}_{ij}) \rangle  \nonumber \\
&&\cong
n(1-3\lambda b)\int {\rm d}\Omega' {\rm d}r' U^{(A)}(r')   g^{(0)}_s[r';(1-3b\lambda)\phi_0 ],   \nonumber\\ \label{pressure_virial3} 
\end{eqnarray}  
where ${\rm d} \Omega'= r'^2\sin\theta {\rm d}\theta{\rm d}\psi$ is the areal element. 
From Eqs. (\ref{pressure_virial}) and (\ref{pressure_virial3}), 
\begin{eqnarray}
&&\dfrac{p}{nT}-1 \cong \nonumber \\
&&\dfrac{\zeta}{6T}n(1-3\lambda b)\int {\rm d}\Omega' {\rm d}r' U^{(A)}(r')   g^{(0)}_s[r';(1-3b\lambda)\phi_0 ]=  \nonumber \\
&&\dfrac{4\zeta }{T} \dfrac{n\pi}{6} [(1-\lambda b)\sigma]^3\int {\rm d}s \biggr(\dfrac{1}{s}\biggr)^\zeta   g^{(0)}_s\biggl\{\sigma s;\dfrac{n\pi}{6}[(1-b\lambda)\sigma]^3 \biggr\}. \nonumber
\\
\label{pressure2} 
\end{eqnarray}  
In the last line $r'$ is replaced by $\sigma s$.  
Equation (\ref{pressure2}) corresponds to the equilibrium pressure of the $A$-system 
with the reduced particle size $(1-\lambda b)\sigma$ 
at a volume fraction of $(1-3\lambda b)\phi_0$ and a number density of $n$.

\section{}

In typical simulation studies, the $\alpha$-relaxation time is identified as the relaxation time of the self-part of the intermediate scattering function. Instead, in this study, the $\alpha$-relaxation time is defined as the relaxation time of the shear-stress autocorrelation function.

\subsection{$\tau_\alpha$: The $A$-system at equilibrium} 
For the $A$-system at equilibrium, the $\alpha\beta$-component of the shear stress ($\alpha\ne \beta$), $\sigma_{\alpha\beta}$, is given as  \cite{Hansen-McdonaldB}
\begin{eqnarray}
\sigma_{\alpha\beta} 
= -\dfrac{1}{2}\sum_i^{N}\sum_{j \ne i} \langle {R}_{\alpha, ij}  \dfrac{\partial}{\partial {R}_{\beta,ij}}U^{(A)}\rangle   \label{stressA}
\end{eqnarray}  
The shear-stress autocorrelation function is $H(t)=\langle {\sigma}_{\alpha\beta}(t){\sigma}_{\alpha\beta}(0)\rangle/L^{3}T$, where $L$ is the system size.    
The $\alpha$-relaxation time $\tau_\alpha^{\rm (eq)}$ is determined by fitting the long-term behavior of $H(t)$ to the Kohlrausch-Williams-Watts (KWW) form $G_0 \exp[-(t/\tau_\alpha)^\chi]$, where $G_0$ is the plateau modulus and $\chi$ is the exponent of nonexponential decay. 

\subsection{${\hat \tau}_\alpha$: The sheared $A$-system}
In the steady state of the $A$-system under the shear flow, Eq. (\ref{shear-flow}), there arises a nonzero average shear stress $\langle \sigma_{xy} \rangle=\langle \sigma_{yx} \rangle$. The deviatoric part of the shear stress is  
\begin{eqnarray}
\delta\sigma_{\alpha\beta} =  \sigma_{\alpha\beta} - \langle \sigma_{xy} \rangle (\delta_{\alpha x}\delta_{\beta y}+\delta_{\alpha y}\delta_{\beta x}), 
\end{eqnarray}
where $\delta_{\alpha\beta}$ is the Kronecker delta. 
Analogous to the determination of the equilibrium $\tau_\alpha$, ${\hat \tau}_\alpha$ is determined by fitting the long-term behavior of the shear-stress autocorrelation function to the KWW form. 

\subsection{${\tilde \tau}_\alpha$: The $B$-system at equilibrium}

For the $B$-system at equilibrium (without shear flow), the $\alpha\beta$-component of the shear stress ($\alpha\ne \beta$), $\sigma_{\alpha\beta}$, is also given as  
\begin{eqnarray}
\sigma_{\alpha\beta} 
= -\dfrac{1}{2}\sum_i^{N}\sum_{j \ne i} \langle {R}_{\alpha ij}  \dfrac{\partial}{\partial {R}_{\beta,ij}}U^{(B)}\rangle. 
\end{eqnarray}  
Similarly, fitting the long-term behavior of the shear-stress autocorrelation function to the KWW form determines ${\tilde \tau}_\alpha$. 
In this case, since the external flow is absent, $\langle \sigma_{\alpha\beta} \rangle=0$. The relaxation time ${\tilde \tau}_\alpha$ is determined by the autocorrelation function of $\sigma_{\alpha\beta}(t)$, and its value is almost the same for any off-diagonal component. 
As noted in the main text, in the $B$-system, because the off-diagonal components of the stress tensor are not symmetric, the net torque is not exactly 0. 
However, the particle configurations are distorted (not due to the external field) so that the resultant local torques are sufficiently suppressed.


\begin{thebibliography}{70}
\bibitem{comment_example}
For example, the two-body correlator along ${\hat x}{\hat y}=1/2$ at $b=-1/2$ is given by $g^{(0)}[r/(1+\lambda),\phi_0(1+3\lambda/2)]$, which is identical with the two-body correlator at $b=1/2$ along the same direction (${\hat x}{\hat y}=1/2$), $g^{(0)}[r,\phi_0(1-3\lambda/2)]$. This is also the case for any $b$ and ${\hat x}{\hat y}$.
\bibitem{comment_cg}
As $\dot\gamma$ increases and $\lambda=c_g\dot\gamma {\hat\tau}_\alpha$ exceeds approximately 0.02, slight deviations from Eq. (28) appear (but they are not very noticeable). 
Although Eq. (28) even seems to be valid, with increasing $\dot\gamma$, $c_g$ gradually decreases, and the isotropic part $g^{(0)}_{s,\mu\nu}(r;\phi_0)$ starts to  exhibit slight $\dot\gamma$ dependence. 
Because of the steep repulsive potential used in the present simulations, increasing $\dot\gamma$ (and thus $\lambda$) to a certain extent may make the growth of particle overlap along the compression axis overwhelming, resulting in that gradual deviation of $g_{s,\mu\nu}({\mbox{\boldmath$r$}};\phi_0)$ from Eq. (28) (with a decrease in $c_g$).
\end{thebibliography}
\end{document}